\documentclass[aps,prb, reprint, citeautoscript, showpacs,superscriptaddress]{revtex4-2}

\usepackage{bm}
\usepackage{amssymb}
\usepackage{graphicx}
\usepackage{amsmath}

\usepackage{textcomp}
\usepackage{colordvi}
\usepackage{multirow}
\usepackage{ulem}
\usepackage{tikz}
\usepackage{color}
\usepackage{makecell}
\usepackage{booktabs}
\usepackage{braket}
\usepackage{hyperref}
\usepackage{subfigure}
\usepackage{float}
\usepackage[figuresright]{rotating}
\usepackage{comment}

\begin{document}
	\title{Photogalvanic effect in few layer graphene}

    \author{Zhaohang Li}
    \affiliation{State Key Laboratory of Integrated Optoelectronics, Key Laboratory of UV-Emitting Materials and Technology of Ministry of Education, Northeast Normal University, Changchun 130024, China.}
    \affiliation{These authors contributed equally to this work.}

	\author{Kainan Chang}
    \affiliation{GPL photonics laboratory, State Key Laboratory of Luminescence Science and Technology, Changchun Institute of Optics, Fine Mechanics and Physics, Chinese Academy of Sciences, Changchun 130033, China.}
    \affiliation{Department of Physics, University of Science and Technology Beijing, Beijing, 100083, China.} 
    \affiliation{These authors contributed equally to this work.}

    \author{Haoyu Li}
    \affiliation{GPL photonics laboratory, State Key Laboratory of Luminescence Science and Technology, Changchun Institute of Optics, Fine Mechanics and Physics, Chinese Academy of Sciences, Changchun 130033, China.}

    \author{Yuxuan Gao}
    \affiliation{State Key Laboratory of Integrated Optoelectronics, Key Laboratory of UV-Emitting Materials and Technology of Ministry of Education, Northeast Normal University, Changchun 130024, China.}

    \author{Wei Xin}
    \email{xinwei@nenu.edu.cn}
    \affiliation{State Key Laboratory of Integrated Optoelectronics, Key Laboratory of UV-Emitting Materials and Technology of Ministry of Education, Northeast Normal University, Changchun 130024, China.}

    \author{Jin Luo Cheng}
    \email{jinluocheng.phys@gmail.com}
    \affiliation{GPL photonics laboratory, State Key Laboratory of Luminescence Science and Technology, Changchun Institute of Optics, Fine Mechanics and Physics, Chinese Academy of Sciences, Changchun 130033, China.}
	\affiliation{School of Physics and Laboratory of Zhongyuan Light, Zhengzhou University, Zhengzhou, China}
	
    \author{Haiyang Xu}
    \email{hyxu@nenu.edu.cn}
    \affiliation{State Key Laboratory of Integrated Optoelectronics, Key Laboratory of UV-Emitting Materials and Technology of Ministry of Education, Northeast Normal University, Changchun 130024, China.}
    
    \date{\today}
	
	\begin{abstract}
		We systematically investigate the nonlinear photogalvanic effect in few-layer graphene with various stacking orders, including AA- and AB-stacked bilayers, and AAA-, ABA-, and ABC-stacked trilayers. 
		 Using a tight-binding model to describe the electronic states, the shift current conductivity and jerk current conductivity are calculated  over a broad spectral range from terahertz to visible  frequencies.
	Our symmetry analysis reveals that a nonvanishing shift current emerges only in ABA-stacked trilayer graphene due to its broken inversion symmetry, with a peak conductivity  reaching approximately $1.21 \times 10^{-13}$ A$\cdot$m/V$^2$ at optimal doping. 
	In contrast, the jerk current, permitted in all structures, requires an in-plane static electric field and exhibits pronounced spectral tunability with chemical potential.
	These findings establish a comprehensive symmetry--band--field coupling paradigm for nonlinear photocurrents in layered graphene and provide design principles for tunable, polarization-sensitive photodetection and energy-harvesting devices based on van der Waals heterostructures.

	\end{abstract}
\maketitle

\section{Introduction}
\label{introduction}
Photogalvanic effects (PGEs) refer to photocurrent phenomena in which direct current (dc) arises from nonlinear light-matter interactions \cite{Cook2017,Sauer2023}.
The direction and magnitude of these resulting photocurrents are determined by the crystal symmetry, electronic band structure, and characteristics of the incident light field, and can be tuned through external environmental conditions \cite{Zheng2023,Zheng2022}. 
Unlike the conventional photoelectric conversion processes that typically rely on built-in electric fields or applied biases, PGEs offer distinct advantages. 
These include a photoresponse not limited by the material's bandgap, the potential to exceed the Shockley-Queisser efficiency limit \cite{Osterhoudt2019, Shockley1961}, and high signal-noise ratio photodetection due to the absence of dark current \cite{Sipe2000}. 
These breakthrough capabilities enable the materials with PGEs as transformative platforms for next-generation optoelectronic devices, particularly in applications such as self-powered photodetection, polarization-sensitive photoelectric sensing, and broadband energy harvesting with unprecedented conversion efficiencies \cite{Postlewaite2024,Novoselov2004}.

Based on the physical mechanisms,  photogalvanic currents can be classified into several categories, such as injection currents, shift currents, jerk currents, drag currents, and nonlinear Hall currents \cite{Sipe2000,Nastos2006,Zheng2022}. 
Among these, the extensively studied shift current, which arises from a second-order nonlinear optical response also known as bulk photovoltaic effect (BPVE), exemplifies symmetry-dependent photoconduction. 
The shift current is inherently restricted to crystal lattices lacking inversion symmetry \cite{Boyd2008,Young2012}.
For centrosymmetric materials, various strategies can be employed to break the inversion symmetry, including applying an out-of-plane electric field, engineering asymmetric band alignments through heterostructure architectures \cite{Postlewaite2024, Akamatsu2021}, utilizing the topological surface states, or leveraging interactions involving electric quadrupoles and magnetic dipoles.
In contrast, the jerk current, which arises from third-order nonlinear optical processes, is not constrained by crystal symmetry and thus exhibits broader applicability across diverse material classes \cite{Muniz2014}. 
Although it requires an external electrostatic driving field, which allows straightforward modulation especially for the transverse current responses, the resulting current amplitude remains inherently limited due to the higher-order nature of the nonlinear response \cite{Fregoso2019}.
This trade-off between generality and output performance highlights the importance of co-designing materials and device architectures for specific applications.

Initial research on the aforementioned PGEs was primarily focused on bulk ferroelectric or topological material systems \cite{Young2012,Muniz2014}.
Recently, however, the focus has shifted towards nanoscale material platforms \cite{Xin2024} considering the growing demand for enhanced photoelectric conversion efficiency and device miniaturization. 
Particularly, two-dimensional (2D) materials have emerged as pivotal candidates due to their unique atomic lattice configurations, tunable band structures, and exceptional optoelectronic properties \cite{Bonaccorso2010,Guha2025}. 
Graphene, a typical 2D material, offers compelling advantages for photocurrent studies due to its ultrahigh carrier mobility, broadband spectral response, gate-tunable optoelectronic behavior, and well-established fabrication and integration protocols \cite{Novoselov2004,Falorsi2025}. 
In 2016,  Pereira’s group pioneered the theoretical description of nonlinear photocurrents in monolayer and bilayer graphene systems using density matrix perturbation expansion, predicting achievable current densities as high as nA$\cdot$cm/W in the visible-to-infrared spectral range \cite{Hipolito2016}.
Subsequently,  Cheng’s group conducted further in-depth studies on the mechanisms and processes of nonlinear photocurrent generation in graphene nanoribbons, bilayer graphene, and twisted bilayer graphene systems \cite{Cheng2018,Zheng2023,Zheng2022}. 
In 2024, Postlewaite's group applied an external electric field to few-layer graphene, enabling it to generate displacement current \cite{Postlewaite2024}. 
More recently, Mao's team explored the shift current in twisted trilayer graphene \cite{Mao2025}.
However, in few-layer graphene systems, the interlayer dependence  of photocurrent generation remains elusive. 
A detailed understanding of the photocurrent generation mechanisms holds important guiding significance for the practical implementation of graphene-based materials in next-generation optoelectronic devices.

To this end,  we investigate the shift and jerk conductivities in bilayer (AA-stacked, AB-stacked) and trilayer (AAA-stacked, ABA-stacked, ABC-stacked) graphene. 
Using tight-binding models for electronic states and perturbative conductivities solved from equation of motion,
we focus on the generation mechanisms and chemical potential dependence.
Notably, among all systems studied, only ABA-stacked trilayer graphene exhibits a significant shift current due to its lack of inversion symmetry, whereas the jerk current, as a universal third-order nonlinear effect, exists in all systems. 
Furthermore, because of the special interlayer coupling in AA-stacked bilayer and AAA-stacked trilayer graphene, their photocurrent generation mechanism is similar to that of a monolayer graphene. 
In contrast, the results of AB-stacked bilayer, ABA-stacked, and ABC-stacked trilayer graphene show totally different carrier transition pathways due to the complex interlayer coupling. 
In addition, we also find that the nonlinear photocurrent  is highly tunable by the chemical potential, with the dominant response wavelength ranging from the terahertz (THz) to the near-infrared regime. 
These findings not only advance fundamental understanding of nonlinear PGE processes in graphene systems, but also provide critical design principles and new avenues for developing tunable optoelectronic devices.

The paper is organized as follows. 
In Sec.\,\ref{Sec2}, we introduce the Hamiltonian of few-layer graphene and the analytical expressions of shift and jerk current conductivities. 
Section\,\ref{Sec3} presents the results and discussion for the shift current in ABA-stacked trilayer graphene,
while Sec.\,\ref{Sec4} shows those of the jerk current in all considered few-layer graphene, with their variations affected by chemical potentials. 
In Sec.\,\ref{Sec5}, we provide estimations for the photocurrent under different light polarizations and photon energies.
Finally, we conclude in Sec.\,\ref{Sec6}.

\section{Model}
\label{Sec2}

\subsection{Electronic Hamiltonian model}

\begin{sidewaystable*}[!ht]
    \renewcommand\arraystretch{1.5}
    \vspace{9cm}    
    \caption{Symmetry and band structure of few-layer graphene. The labels A and B represent the two inequivalent carbon atoms. The electronic couplings are parameterized by: $\gamma_{0}$ for the intralayer nearest-neighbor hopping, $\gamma_{1}$ for the interlayer nearest-neighbor hopping, $\gamma_{3}$ for the interlayer second-nearest-neighbor hopping.
    For ABA-TG, $\gamma_{2}$ and $\gamma_{5}$ represent the coupling between the top and bottom layers without and with other atoms in between, respectively. The lattice constant $a_{0} = 2.46$ \AA , layer spacing $c = 3.35$ \AA .
    For AB-TG, $\Delta'$ represent the on-site potential difference induced by the asymmetric environment of the A and B atoms in the crystal structure.
    Lattice vectors are taken as $\boldsymbol{a}_{1} = \left(\frac{1}{2}, \frac{\sqrt{3}}{2}\right) a_{0}$ and $\boldsymbol{a}_{2} = \left(-\frac{1}{2}, \frac{\sqrt{3}}{2}\right) a_{0}$. The reciprocal lattice vectors are $\boldsymbol{b}_{1} = \frac{2\pi}{a_{0}}\left(\hat{\boldsymbol{x}} + \frac{1}{\sqrt{3}} \hat{\boldsymbol{y}} \right)$ and $\boldsymbol{b}_{2} = \frac{2\pi}{a_{0}}\left(-\hat{\boldsymbol{x}} + \frac{1}{\sqrt{3}} \hat{\boldsymbol{y}} \right)$. $\hat{\boldsymbol{x}}$, $\hat{\boldsymbol{y}}$ and $\hat{\boldsymbol{z}}$ are unit vector. $g_{0}=1+e^{-i\boldsymbol{k}\cdot \boldsymbol{a}_{1}}+e^{-i\boldsymbol{k}\cdot \boldsymbol{a}_{2}}$, $g_{1}=e^{-i\boldsymbol{k}\cdot \boldsymbol{a}_{1}}+e^{-i\boldsymbol{k}\cdot \boldsymbol{a}_{2}}+e^{-i\boldsymbol{k}\cdot (\boldsymbol{a}_{1} + \boldsymbol{a}_{2})}$, $g_{2}=e^{-i\boldsymbol{k}\cdot (\boldsymbol{a}_{1} + \boldsymbol{a}_{2})}$ \cite{Chuang2012,Brun2015,Cortes2021}.}
    \begin{tabular}{cccccccc} 
        \hline\hline
        \multirow{2}{*}{Type}& \multirow{2}{*}{\shortstack{Structural \\ diagram}} & \multirow{2}{*}{\shortstack{Coupling\\ parameter\\ (eV)}} & \multirow{2}{*}{\shortstack{Point \\ group}} & \multirow{2}{*}{\shortstack{Atom \\position}} & \multirow{2}{*}{Hamiltonian} &  \multicolumn{2}{c}{Non-zero tensor elements of conductivity} \\
        \cline{7-8}
        &  &  &  &  & & $\sigma^{(2)}$ & $\sigma^{(3)}$ \\\hline
        AA-BG & \raisebox{-.5\height} {\includegraphics[width=40pt]{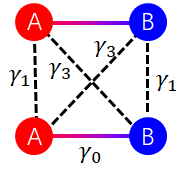} } & 
        $\begin{matrix}
            \gamma_{0}=2.569  \\
            \gamma_{1}=0.361 \\
            \gamma_{3}=-0.032 \\
        \end{matrix}$ &
        $D_{6h}$ &
        $\begin{matrix}
            \boldsymbol{\tau}_{A}^{1} = \boldsymbol{0} \\
    	    \boldsymbol{\tau}_{B}^{1} = \left(\boldsymbol{a}_{1} + \boldsymbol{a}_{2}\right)/3 \\
    	    \boldsymbol{\tau}_{A}^{2} = c\hat{\boldsymbol{z}} \\
    	    \boldsymbol{\tau}_{B}^{2} = \boldsymbol{\tau}_{B}^{1} + c\hat{\boldsymbol{z}}
        \end{matrix}$ & 
        $\left(\setlength{\arraycolsep}{5pt}    
         \renewcommand{\arraystretch}{1} 
         \begin{array}{cccc}
            0 & \gamma_{0}g_{0} & \gamma_{1} & \gamma_{3}g_{0} \\
            \gamma_{0}g_{0}^{*} & 0 & \gamma_{3}g_{0}^{*} & \gamma_{1} \\
            \gamma_{1} & \gamma_{3}g_{0} & 0 & \gamma_{0}g_{0} \\
            \gamma_{3}g_{0}^{*} & \gamma_{1} & \gamma_{0}g_{0}^{*} & 0
\end{array} \right)$ & 
        \multirow{2}{*}[-1cm]{\makecell[c]{Each\\ element\\ vanishes}} & 
        \multirow{3}{*}[-2cm]{\makecell[c]{$zzzz$, \\$xxxx=yyyy$\\$=xxyy+xyyx+xyxy$, \\$\left\{ \begin{aligned}
    	    xxyy=yyxx\\
    	    xyyx=yxxy\\
    	    xyxy=yxyx
        \end{aligned}\right.$, \\$yyzz = xxzz$, $zzyy = zzxx$, \\$zyyz = zxxz$, $yzzy = xzzx$, \\$yzyz = xzxz$, $zyzy = zxzx$}} \\
        \cline{1-6} 
        AAA-TG & \raisebox{-.5\height}{\includegraphics[width=40pt]{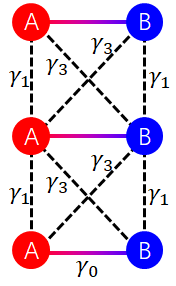}} &
        $\begin{matrix}
    	\gamma_{0}=3.2  \\
    	\gamma_{1}=0.18  \\
        \end{matrix}$ &
        $D_{6h}$ &
        $\begin{matrix}
    	\boldsymbol{\tau}_{A}^{1} = \boldsymbol{0} \\
    	\boldsymbol{\tau}_{B}^{1} = \left(\boldsymbol{a}_{1} + \boldsymbol{a}_{2} \right)/3 \\
    	\boldsymbol{\tau}_{A}^{2} = c\hat{\boldsymbol{z}} \\
    	\boldsymbol{\tau}_{B}^{2} = \boldsymbol{\tau}_{B}^{1} + c\hat{\boldsymbol{z}} \\
    	\boldsymbol{\tau}_{A}^{3} = 2c\hat{\boldsymbol{z}} \\
    	\boldsymbol{\tau}_{B}^{3} = \boldsymbol{\tau}_{B}^{1} + 2c\hat{\boldsymbol{z}}
        \end{matrix}$ &
                $\left(\setlength{\arraycolsep}{2pt}    
         \renewcommand{\arraystretch}{1} 
         \begin{array}{cccccc}
    	0 & \gamma_{0}g_{0} & \gamma_{1} & 0 & 0 & 0 \\
    	\gamma_{0}g_{0}^{*} & 0 & 0 & \gamma_{1} & 0 & 0 \\
    	\gamma_{1} & 0 & 0 & \gamma_{0}g_{0} & \gamma_{1} & 0 \\
    	0 & \gamma_{1} & \gamma_{0}g_{0}^{*} & 0 & 0 & \gamma_{1} \\
    	0 & 0 & \gamma_{1} & 0 & 0 & \gamma_{0}g_{0} \\
    	0 & 0 & 0 & \gamma_{1} & \gamma_{0}g_{0}^{*} & 0 \\
        \end{array}\right)$ & & \\
        \cline{1-7} 
        ABA-TG & \raisebox{-.5\height}{\includegraphics[width=60pt]{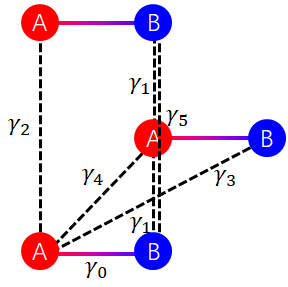}} &
        $\begin{matrix}
    	\gamma_{0}=3.16  \\
    	\gamma_{1}=0.39  \\
    	\gamma_{2}=-0.02  \\
    	\gamma_{3}=0.315  \\
    	\gamma_{4}=0.044  \\
    	\gamma_{5}=-0.04
        \end{matrix}$ &
        $D_{3h}$ & 
        $\begin{matrix}
    	\boldsymbol{\tau}_{A}^{1} = \boldsymbol{0} \\
    	\boldsymbol{\tau}_{B}^{1} = \left(\boldsymbol{a}_{1} + \boldsymbol{a}_{2} \right)/3 \\
    	\boldsymbol{\tau}_{A}^{2} = \boldsymbol{\tau}_{B}^{1} + c\hat{\boldsymbol{z}} \\
    	\boldsymbol{\tau}_{B}^{2} = 2\boldsymbol{\tau}_{B}^{1} + c\hat{\boldsymbol{z}} \\
    	\boldsymbol{\tau}_{A}^{3} = 2c\hat{\boldsymbol{z}} \\
    	\boldsymbol{\tau}_{B}^{3} = \boldsymbol{\tau}_{B}^{1} + 2c\hat{\boldsymbol{z}}
        \end{matrix}$ &
        $\left(\setlength{\arraycolsep}{2pt}    
         \renewcommand{\arraystretch}{1} 
         \begin{array}{cccccc}
    	0 & \gamma_{0}g_{0} & \gamma_{4}g_{0} & \gamma_{3}g_{1} & \gamma_{2} & 0 \\
    	\gamma_{0}g_{0}^{*} & 0 & \gamma_{1} & \gamma_{4}g_{0} & 0 & \gamma_{5} \\
    	\gamma_{4}g_{0}^{*} & \gamma_{1} & 0 & \gamma_{0}g_{0} & \gamma_{4}g_{0}^{*} & \gamma_{1} \\
    	\gamma_{3}g_{1}^{*} & \gamma_{4}g_{0}^{*} & \gamma_{0}g_{0}^{*} & 0 & \gamma_{3}g_{1}^{*} & \gamma_{4}g_{0}^{*} \\
    	\gamma_{2} & 0 & \gamma_{4}g_{0} & \gamma_{3}g_{1} & 0 & \gamma_{0}g_{0} \\
    	0 & \gamma_{5} & \gamma_{1} & \gamma_{4}g_{0} & \gamma_{0}g_{0}^{*} & 0 \\
        \end{array}\right)$ &
        $\begin{matrix}
    	-xxy \\ = -xyx \\ = -yxx \\ = yyy
        \end{matrix}$ & \\
        \hline
        AB-BG & \raisebox{-.5\height}{ \includegraphics[width=60pt]{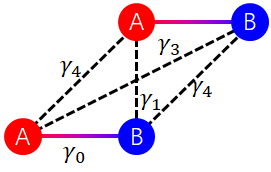}} &
        $\begin{matrix}
        \gamma_{0}=3.16  \\
        \gamma_{1}=0.381  \\
        \gamma_{3}=0.38  \\
        \gamma_{4}=0.14  \\
        \Delta^\prime = 0.022 \\
        \end{matrix}$ &
        $D_{3d}$ &
        $\begin{matrix}
    	\boldsymbol{\tau}_{A}^{1} = \boldsymbol{0} \\
    	\boldsymbol{\tau}_{B}^{1} = \left(\boldsymbol{a}_{1} + \boldsymbol{a}_{2} \right)/3 
    	\\
    	\boldsymbol{\tau}_{A}^{2} = \boldsymbol{\tau}_{B}^{1} + c\hat{\boldsymbol{z}} \\
    	\boldsymbol{\tau}_{B}^{2} = 2\boldsymbol{\tau}_{B}^{1} + c\hat{\boldsymbol{z}}
        \end{matrix}$ &
                $\left(\setlength{\arraycolsep}{5pt}    
         \renewcommand{\arraystretch}{1} 
         \begin{array}{cccc}
         -\frac{\Delta'}{2} & -\gamma_{0}g_{0} & \gamma_{4}g_{0} & -\gamma_{3}g_{0}^{*} \\
        -\gamma_{0}g_{0}^{*} & \frac{\Delta'}{2} & \gamma_{1} & \gamma_{4}g_{0} \\
        \gamma_{4}g_{0}^{*} & \gamma_{1} & \frac{\Delta'}{2} & -\gamma_{0}g_{0} \\
        -\gamma_{3}g_{0} & \gamma_{4}g_{0}^{*} & -\gamma_{0}g_{0}^{*} & -\frac{\Delta'}{2}
        \end{array}\right)$ &
        \multirow{2}{*}[-1.cm]{\makecell[c]{Each \\element \\vanishes}} &
        \multirow{2}{*}[0.5cm]{{\makecell[c]{$zzzz$, \\$xxxx=yyyy$\\$=xxyy+xyyx+xyxy$, \\$\left\{ \begin{aligned}
    			xxyy=yyxx\\
    			xyyx=yxxy\\
    			xyxy=yxyx
    		\end{aligned}\right.$, \\$yyzz = xxzz$, $zzyy = zzxx$, \\$zyyz = zxxz$, $yzzy = xzzx$, \\$yzyz = xzxz$, $zyzy = zxzx.$, \\$xxxz = -xyyz = -yxyz = -yyxz$, \\$xxzx = -xyzy = -yxzy = -yyzy$, \\$xzxx = -xzyy = -yzxy = -yzyx$, \\$zxxx = -zxyy = -zyxy = -zyyx$}}} \\
        \cline{1-6}
        ABC-TG & \raisebox{-.5\height}{\includegraphics[width=70pt]{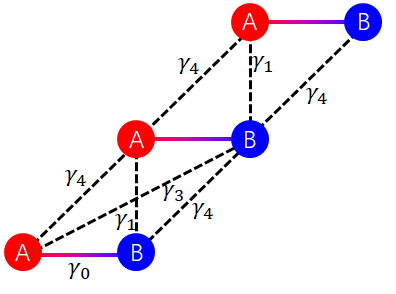}} &
        $\begin{matrix}
    	\gamma_{0}=3.1  \\
    	\gamma_{1}=0.4 \\
    	\gamma_{3}=0.2 \\
        \end{matrix}$ &
        $D_{3d}$ & 
        $\begin{matrix}
    	\boldsymbol{\tau}_{A}^{1} = \boldsymbol{0} \\
    	\boldsymbol{\tau}_{B}^{1} = \left(\boldsymbol{a}_{1} + \boldsymbol{a}_{2} \right)/3 \\
    	\boldsymbol{\tau}_{A}^{2} = \boldsymbol{\tau}_{B}^{1} + c\hat{\boldsymbol{z}} \\
    	\boldsymbol{\tau}_{B}^{2} = 2\boldsymbol{\tau}_{B}^{1} + c\hat{\boldsymbol{z}} \\
    	\boldsymbol{\tau}_{A}^{3} = \boldsymbol{\tau}_{B}^{2} + c\hat{\boldsymbol{z}} \\
    	\boldsymbol{\tau}_{B}^{3} = 3\boldsymbol{\tau}_{B}^{1} + 2c\hat{\boldsymbol{z}}
        \end{matrix}$ &
                $\left(\setlength{\arraycolsep}{2pt}    
         \renewcommand{\arraystretch}{1} 
         \begin{array}{cccccc}
    	0 & \gamma_{0}g_{0} & \gamma_{4}g_{0} & \gamma_{3}g_{1} & 0 & \gamma_{2}g_{2} \\
    	\gamma_{0}g_{0}^{*} & 0 & \gamma_{1} & \gamma_{4}g_{0} & 0 & 0 \\
    	\gamma_{4}g_{0}^{*} & \gamma_{1} & 0 & \gamma_{0}g_{0} & \gamma_{4}g_{0} & \gamma_{3}g_{1} \\
    	\gamma_{3}g_{1}^{*} & \gamma_{4}g_{0}^{*} & \gamma_{0}g_{0}^{*} & 0 & \gamma_{1} & \gamma_{4}g_{0} \\
    	0 & 0 & \gamma_{4}g_{0}^{*} & \gamma_{1} & 0 & \gamma_{0}g_{0} \\
    	\gamma_{2}g_{2}^{*} & 0 & \gamma_{3}g_{1}^{*} & \gamma_{4}g_{0}^{*} & \gamma_{0}g_{0}^{*} & 0 \\
        \end{array}\right)$  & & \\
        \hline
        \hline
    \end{tabular}  
    \label{Tab1}   
\end{sidewaystable*}

This study focuses on bilayer and trilayer graphene with different interlayer stacking orders, including AA- and AB-stacked bilayer, along with AAA-, ABA-, and ABC-stacked trilayer (denoted as AA-BG, AB-BG, AAA-TG, ABA-TG, and ABC-TG, 
respectively, as listed in Table\,\ref{Tab1}). 
These crystal structures possess distinct point group symmetries, leading to structure-dependent Hamiltonian and  associated physical properties \cite{Boyd2008}.
For each stacking configuration, the Hamiltonian $\hat{H}_{k}$ is constructed by the tight-binding model, formed by the carbon $p_z$ orbitals.
Table\,\ref{Tab1} summarizes the detailed information of each system,
including lattice constants, interlayer distances, primitive vectors, 
atomic positions per unit cell,  
interlayer/intralayer coupling parameters ($\gamma$), 
and 
structure factors (wavevector $\bm k$ dependent functions $g_0(\bm k)$, $g_1(\bm k)$, and $g_2(\bm k)$) \cite{Boyd2008,Zheng2022,Zheng2023,Chuang2012,Brun2015,Cortes2021}. .
Note, for AB-BG, due to the asymmetry between A and B atoms, the on-site energy difference ($\Delta^\prime$) is also considered. %
After diagonalizing each Hamiltonian,
\begin{align}
    \hat{H}_{\boldsymbol{k}}C_{n\boldsymbol{k}} = \epsilon_{n\boldsymbol{k}}C_{n\boldsymbol{k}}\,,
\end{align}
we obtain the eigenvectors $C_{n\boldsymbol{k}}$ and the eigenenergies $\epsilon_{n\boldsymbol{k}}$ for the $n$-th band at the wavevector $\boldsymbol{k}$.
The computation of nonlinear photocurrent conductivities relies on the position operator ($\hat{\bm{r}}_{\bm{k}}$) and the velocity operator ($\hat{\bm{v}}_{\bm{k}}$).
In the tight-binding Bloch basis, the matrix elements of the position operator are
\begin{align}	
	{\bm{r}}_{\alpha_1 \alpha_2\bm{k}} = i \nabla_{\bm{k}} \delta_{\alpha_1 \alpha_2}+ \bm{\tau}_{\alpha_1} \delta_{\alpha_1 \alpha_2}\,,
\end{align}
and those of the velocity operator are
\begin{align}	
\bm{v}_{n_1n_2\bm{k}} = C^\dagger_{n_1\bm{k}} \hat{\bm{v}}_{\bm{k}} C_{n_2\bm{k}}\,,
\label{v}
\end{align}
with
\begin{align}	
\hat{\bm{v}}_{\bm{k}} = \frac{1}{i\hbar} \left[\hat{\bm{r}}_{\bm{k}}, \hat{H}_{\bm{k}}\right]\,.
\end{align}
Similarly, we also express the Berry connection in the eigenstate basis as
\begin{align}
	\bm{\xi}_{n_1n_2\bm{k}} = C^\dagger_{n_1\bm{k}} \hat{\bm{r}}_{\bm{k}} C_{n_2\bm{k}}\,.
    \label{berry}
\end{align}
The off-diagonal components of the Berry connection can be derived from the velocity matrix elements
\begin{align}
	\bm{r}_{n_1n_2\bm{k}} = 
	\begin{cases} 
		\bm{\xi}_{n_1n_2\bm{k}} = \frac{\bm{v}_{n_1n_2\bm{k}}}{i\omega_{n_1n_2\bm{k}}} & \text{for } n_1 \neq n_2 \\
		0 & \text{for } n_1 = n_2
	\end{cases}\,,
	\label{r}
\end{align}
where $\hbar\omega_{n_1n_2\bm{k}} = \epsilon_{n_1\bm{k}} - \epsilon_{n_2\bm{k}}$.
 For completeness, we note that $\bm{\xi}_{n_1n_1\bm{k}}$ is not directly obtainable from this relation and must be handled separately.
The diagonal components $\xi^{a}_{n_1n_1\bm{k}}$ appear in the covariant derivative
\begin{equation}
	(r^{c}_{\bm{k}})_{;n_1n_2k^{a}} = \frac{\partial r^{c}_{n_1n_2\bm{k}}}{\partial k^{a}} - i(\xi^{a}_{n_1n_1\bm{k}} - \xi^{a}_{n_2n_2\bm{k}})r^{c}_{n_1n_2\bm{k}}\,.
\end{equation}
This can be alternatively evaluated as:
\begin{equation}
	(r^{c}_{\bm{k}})_{;n_1n_2k^{a}} = \frac{-ir^{c}_{n_1n_2\bm{k}}{\Delta}^{a}_{n_2n_1\bm{k}} + \hbar {M}^{ca}_{n_1n_2\bm{k}} + i[r^{a}_{\bm{k}}, v^{c}_{\bm{k}}]_{n_1n_2}}{i\omega_{n_1n_2\bm{k}}}\,,
\end{equation}
with ${\Delta}^{a}_{n_2n_1\bm{k}} = v^{a}_{n_2n_2\bm{k}} - v^{a}_{n_1n_1\bm{k}} = \frac{\partial\omega_{n_2n_1\bm{k}}}{\partial k^{a}}$ and
\begin{equation}
	{M}^{ca}_{n_1n_2\bm{k}} = {C}^{\dagger}_{n_1\bm{k}}\frac{1}{i\hbar}[\hat{r}^{a}_{\bm{k}}, \hat{v}^{c}_{\bm{k}}] {C}_{n_2\bm{k}}\,.
\end{equation}
The wavevector is confined to the in-plane directions ($x$ and $y$). Consequently, the out-plane covariant derivative simplifies to
\begin{equation}
	(r^{a}_{k})_{;n_1n_2k^{z}} = -i(\xi^{z}_{n_1n_1k} - \xi^{z}_{n_2n_2k})r^{a}_{n_1n_2k}\,.
\end{equation}

\subsection{Conductivity expression}

The PGE describes the generation of dc from nonlinear optical responses under illumination by an electromagnetic wave $\boldsymbol{E}(t)=\boldsymbol{E}_{\omega}e^{-i\omega t}+c.c.$ with frequency $\omega$ \cite{Chan2017}, in the presence of a static electric field $\boldsymbol{E}_{\text{dc}}$. 
The total dc density can be written as:
\begin{align}
    \boldsymbol{J}_{\text{dc}} = \boldsymbol{J}_{\text{dc}}^{(1)} + \boldsymbol{J}_{\text{dc}}^{(2)} + \boldsymbol{J}_{\text{dc}}^{(3)} + \cdots\,,
\end{align}
where the first term $J_{\text{dc}}^{(1);a} = \sigma^{(1);ab}E_{\text{dc}}^{b}$ is the photocurrent induced by the dc field with the linear conductivity $\sigma^{(1);ab}$, the second term ${J}_{\text{dc}}^{(2);a} = \sigma^{(2);abc}(\omega , -\omega)E_{\omega}^{b}E_{-\omega}^{c} $ corresponds to the shift current characterized by the second-order conductivity $\sigma^{(2);abc}(\omega , -\omega)$, and the third term ${J}_{\text{dc}}^{(3);a} = \sigma^{(3);abcd}(\omega , -\omega , 0)E_{\omega}^{b}E_{-\omega}^{c}E_{\text{dc}}^{d} $ describes the jerk current characterized by the third-order conductivity $\sigma^{(3);abcd}(\omega , -\omega , 0)$. 
These two nonlinear conductivities are the focus of this work, and their spectra can be calculated when band eigenenergies, Berry connections, and velocity matrix elements are known.
The detailed microscopic expressions 
 for these conductivities have been derived in previous works \cite{Chang2026}.

As the nonlinear conductivities are tensors governed by crystal symmetry, a symmetry analysis is essential to determine their independent non-vanishing components. 
Therefore, we begin by defining the crystal orientation for the few-layer graphene systems studied here: the armchair crystal direction as the $x$-axis, and the zigzag crystal direction as the $y$-axis. 
The point groups of these crystals and the nonzero components of the conductivity tensors are listed in Table\,\ref{Tab1}.
The linear conductivity is in-plane isotropic, yielding $\sigma^{(1);xx} = \sigma^{(1);yy} = \sigma_{\text{dc}}$. 
For the second-order conductivity $\sigma^{(2);abc}(\omega, -\omega)$, nonzero components exist only for ABA-TG, satisfying $\sigma^{(2);xxy} = \sigma^{(2);xyx} = \sigma^{(2);yxx} = -\sigma^{(2);yyy}$. 
The third-order conductivity $\sigma^{(3);abcd}(\omega , -\omega , 0)$  has the same nonzero in-plane tensor components, for all the discussed few-layer graphene.
There exist two independent nonzero in-plane components $\sigma^{(3);xxyy}$ and $\sigma^{(3);xyyx}$. All others satisfy $\sigma^{(3);xxxx} = \sigma^{(3);yyyy} = 2\text{Re}[\sigma^{(3);xxyy}] + \sigma^{(3);xyyx} $, and $\sigma^{(3);yxxy} = \sigma^{(3);xyyx}$ and $\sigma^{(3);yyxx} = \sigma^{(3);xxyy} = [\sigma^{(3);yxyx}]^{*} = [\sigma^{(3);xyxy}]^{*} $, among which we have used the relation $\sigma^{(3);abcd}(\omega , -\omega , 0) = [\sigma^{(3);abcd}(-\omega , \omega , 0)]^{*} = [\sigma^{(3);acbd}(\omega , -\omega , 0)]^{*}$ to ensure the physical current density as a real quantity; consequently, $\sigma^{(3);xyyx}$ is real and $\sigma^{(3);xxyy}$ is complex.

For a normally incident light with frequency $\omega$, the electric field can be written as
\begin{align}
    \boldsymbol{E}_{\omega} &= E_\omega^x\hat{\bm x}+E_\omega^y\hat{\bm y}\,.
    \label{field}
\end{align}
 This field at the negative frequency  $-\omega$ is the complex conjugate, $\bm{E}_{-\omega}=\bm{E}_{\omega}^*$. 
Meanwhile, a dc field is applied along an in-plane direction, as
\begin{align}
\bm{E}_{\mathrm{dc}} = E_{\mathrm{dc}}\hat{\bm{e}}_\parallel\,.
\end{align}
With these definitions, the total photocurrent is then given by
\begin{widetext}
\begin{align}
    \boldsymbol{J}_{\text{dc}} = &
    \left[\sigma_{\text{dc}} + 6\sigma^{(3);xyyx}(\omega,-\omega,0)
    \left|\boldsymbol{E}_{\omega}\right|^2
    + 12 \text{Re}[
    \sigma^{(3);xxyy}(\omega,-\omega,0)]
    \left|\boldsymbol{E}_{\omega}\cdot \hat{\boldsymbol{e}}_{\parallel}\right|^2
     \right]
     \bm{E}_{\text{dc}} \notag \\
    &+ 2\sigma^{(2);xxy}(\omega,-\omega)\begin{pmatrix}
        2\text{Re}[E_{\omega}^{x}(E_{\omega}^{y})^*] \\
        \left|E_{\omega}^{x}\right|^{2} - \left|(E_{\omega}^{y})^*\right|^{2}
    \end{pmatrix} \notag \\
    &+ 12 \text{Re} [ \sigma^{(3);xxyy}(\omega, -\omega , 0)
     \boldsymbol{E}_{\omega } \cdot \hat{\boldsymbol{e}}_{\bot} \boldsymbol{E}_{\omega}^* \cdot \bm{E}_{\text{dc}}]\hat{\boldsymbol{e}}_{\bot}\,,
     \label{current}
\end{align}
\end{widetext}
with $\hat{\bm{e}}_\bot=\hat{\bm{z}}\times\hat{\bm{e}}_\parallel$.
On the right-hand side of this equation, the first line denotes the current parallel to $\bm{E}_{\text{dc}}$, which consists of the linear conductivity  $\sigma_{\text{dc}}$ and two third-order jerk conductivities, one associated with $\sigma^{(3);xyyx}(\omega,-\omega,0)$, and the other associated with $\sigma^{(3);xxyy}(\omega,-\omega,0)$; 
the second line represents the shift current with $\sigma^{(2);xxy}(\omega,-\omega)$, which
is independent of the dc field;
and the third line gives the third-order jerk current, which is also associated with $\sigma^{(3);xxyy}(\omega,-\omega,0)$,
 while perpendicular to $\bm{E}_{\text{dc}}$.

\section{Results for shift current in ABA-TG}
\label{Sec3}

\begin{figure*}[!htp]
	\centering
	\includegraphics[width=\linewidth]{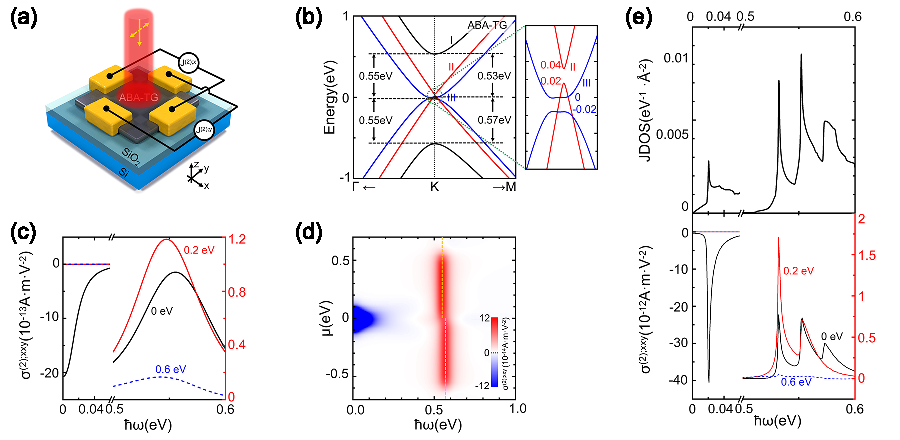}
	\caption{
	(a) Schematic of the shift current response collected by electrodes along the $x$- and $y$-directions.
	(b) Electronic band structure of ABA-TG. Insert is a magnification of the middle part.
	Spectra of the shift conductivity $\sigma^{(2);xxy}$ for ABA-TG at (c) discrete chemical potentials $\mu =0, 0.2, 0.6$ eV and (d) at continuously varying chemical potentials, with $\gamma=33$ meV.
	(e) The lower half  is similar to (c), but with $\gamma=0.5$ meV, and the upper half is JDOS of ABA-TG.
	}
	\label{fig1}
\end{figure*}

In this work, we adopt the following computational parameters: a uniform $3000\times3000$ $\bm k$-grid in Brillouin zone, a temperature of $T=300$ K, and a damping coefficient of $\gamma=33$ meV.
We begin by discussing the shift current in the few-layer graphene.
As established by our symmetry analysis, 
among all structures considered, 
only ABA-TG exhibits a nonzero shift current.
A schematic diagram of the measurement is provided in Fig.\,\ref{fig1}\,(a).

The electronic band structure of ABA-TG within $\pm1$ eV is presented in Fig.\,\ref{fig1}\,(b). 
Three distinct sets of bands can be identified: 
(I) a pair of gapped parabolic bands, separated from the zero energy by approximately 0.53 eV (above) and 0.57 eV (below), respectively;
(II) a pair of linear bands; 
(III) a pair of parabolic bands that intersect the linear bands within a narrow momentum region, as highlighted in the zoomed inset.
We note that the energy separations at K point between the top of lower band in set III and edges of two bands in set I are both 0.55 eV, and in addition, the energy splittings within sets II and III are both 0.02 eV.

The corresponding shift conductivity spectra are displayed in Fig.\,\ref{fig1}\,(c). 
At $\mu=0$ eV, a strong shift current response  is observed at photon energies below 0.06 eV  with an amplitude of approximately $-200 \times 10^{-13}$ A$\cdot$m$\cdot$V$^{-2}$, accompanied by a prominent peak near 0.55 eV.
To elucidate the microscopic origin of these features, we 
recalculate the shift conductivity with smaller of $\gamma=0.5$ meV, which also requires a denser $\bm k$-grid of $12000\times12000$ points. 
The resulting shift conductivity spectra of $\sigma^{(2);xxy}$ are given in Fig.\,\ref{fig1}\,(e).
At $\mu = 0$ eV, the spectrum exhibits a pronounced lower-energy peak at $\hbar\omega=0.02$ eV, along with three obvious peaks at higher energies---specifically  at $\hbar\omega=0.53$, 0.55, and 0.57 eV.
To clarify these features,
the joint density of states (JDOS) is given in Fig.\,\ref{fig1}\,(e), defined as
\begin{align}
\mathrm{JDOS}=
\int \frac{d{\bm k}}{(2\pi)^2}
\sum_{nm}f_{nm{\bm k}} \delta(\omega_{mnk}-\omega)\,.
\end{align}
We can find that there are four distinct peaks at 0.02, 0.53, 0.55, and 0.57 eV.
Based on these energy differences at the K point and the JDOS profile, we can reasonably attribute the three higher-energy conductivity peaks to interband transitions involving sets I and III, and the lower-energy peak originates from combined transitions within set II and/or set III.
Consequently, in Fig.\,\ref{fig1}\,(e)
the lower-energy peak vanishes once $\mu$ exceeds 0.04 eV, as the relevant transitions within sets II and III  are Pauli-blocked.
At $\mu$ = 0.2 eV, the third higher-energy peak near 0.57 eV disappears, due to the occupation of the corresponding initial and final states.
When $\mu$ further raises to 0.6 eV, all interband transitions are suppressed within the considered spectral window, leaving only weak residual features at 0.53 and 0.55 eV.

Returning to Fig.\,\ref{fig1}\,(c), we see that the higher-energy peak under normal damping ($\gamma=33$ meV) merges the three distinct higher-energy peaks identified under the small damping condition; and the lower-energy peak originates from two bands of set III.
As the chemical potential increases to $\mu$ = 0.2 and 0.6 eV, the lower-energy response vanishes;
for $\mu$ = 0.2 eV, 
the higher-energy peak associated with the first two peaks remain observable and exhibits an obvious redshift, while for for $\mu$ = 0.6 eV, all spectral features are suppressed.
A broader range of chemical potentials is explored in Fig.\,\ref{fig1}\,(d). 
It reveals that the merged conductivity features near 0.55 eV (originating from transitions at 0.53 and 0.55 eV)
persist for all $\mu$ below 0.53 eV, and those near 0.57 eV (involving the 0.55 and 0.57 eV transitions)  also remain visible within the respective chemical potential range.
In addition, low-photon-energy shift conductivity ($<0.2$ eV) remains obvious for $|\mu| < 0.1$ eV, which can be attributed to the high density of states at Dirac point enabling significant carrier transitions.



\section{Results for Jerk Current}
\label{Sec4}

The jerk current, as a third-order optical nonlinearity, is not constrained by crystal symmetry and thus exists in all graphene structures. A schematic diagram of this process is shown in Fig.\,\ref{fig2}\,(a).
Based on their band structures, these five graphene structures are classified into two groups as shown in Fig.\,\ref{fig2}\,(b): group I, which includes AA-BG and AAA-TG, exhibits linear dispersion relations inherited from the Dirac cones of monolayer graphene; group II, which includes AB-BG, ABA-TG, and ABC-TG, displays different band features,
such as parabolic bands,
arising from special interlayer coupling.

\subsection{AA-BG and AAA-TG}
\label{Sec41}

\begin{figure*}[htp]
	\centering
	\includegraphics[width=\linewidth]{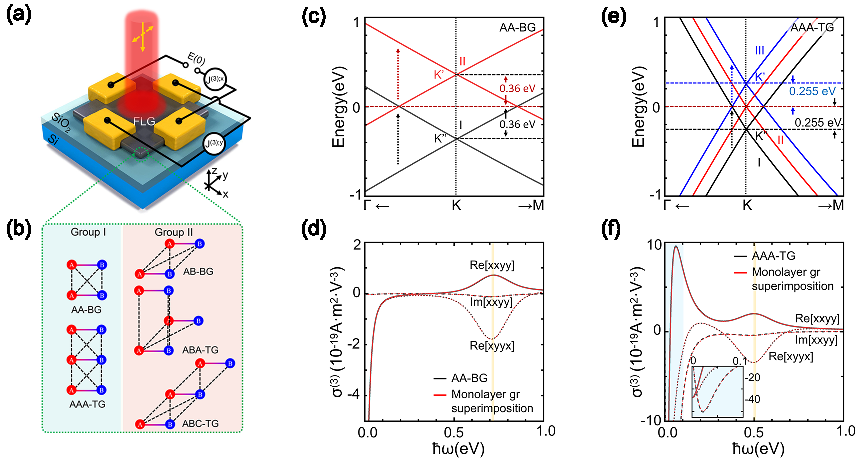}
	\caption{(a) Schematic of the jerk current response collected by electrodes along the $x$- and $y$-directions  under an electrostatic field applied along the $y$-direction.
	(b) Two groups of considered few-layer graphene with structural diagrams.	
	(c) Electronic band structure of AA-BG. 
	(d) Spectra of the jerk conductivities Re[$\sigma^{(3);xxyy}$], Im[$\sigma^{(3);xxyy}$], and Re[$\sigma^{(3);xyyx}$] for AA-BG at a chemical potential $\mu=0$ eV (black lines), and for the combined two layers of monolayer graphene (red lines).
	(e) Electronic band structure of AAA-BG. 
	(f) Spectra of the jerk conductivities Re[$\sigma^{(3);xxyy}$], Im[$\sigma^{(3);xxyy}$], and Re[$\sigma^{(3);xyyx}$] for AAA-BG at a chemical potential $\mu=0$ eV (black lines), and for the combined three layers of monolayer graphene (red lines).
	Note, the superimposed spectra are given by Eqs.\,\eqref{Gr-AA} and \eqref{Gr-AAA}.
	}
	\label{fig2}
\end{figure*}

Figure\,\ref{fig2}\,(c) illustrates the band structure of AA-BG. 
Unlike monolayer graphene with a single Dirac cone, AA-BG exhibits 
two nearly parallel sets of linearly dispersing bands (labeled  I and II, distinguished by red and black lines, respectively). 
Each set intersects  the corner at a distinct K point (labeled K$'$ and K$''$), thereby forming two separate Dirac cones.
These two cones are offset in energy by approximately 0.72 eV.
Due to the absence of strict electron-hole symmetry in the bilayer, their crossing points lie near but not exactly at zero energy.

The jerk conductivity spectra of AA-BG at $\mu=0$ eV, shown as black lines in Fig.\,\ref{fig2}\,(d), exhibit the single resonant peaks around 0.72 eV. 
We can attribute these peaks to the intracone transitions near the Fermi surface, as indicated by the red and black dotted arrows in Fig.\,\ref{fig2}\,(c). 
To further verify this interpretation, we directly compare the AA-BG conductivity with that of monolayer graphene.
Specifically, the jerk conductivities of monolayer graphene are calculated at chemical potentials $\mu=\pm0.361$ eV corresponding to the energies of K$'$ and K$''$ intersecting points.
The simple superpositions of these two monolayer contributions are shown as the red lines in Fig.\,\ref{fig2}\,(d). 
After applying scaling factors that account for the Fermi velocity renormalization between the materials, the combined monolayer spectra show remarkable agreement with those of AA-BG.
Note, the origin of these scaling factors can be traced to the low-energy band structure.
The eigen-energy of AA-BG is expressed by
$\epsilon^{\mathrm{AA}}_{n\bm{k}} 
= \alpha^{\mathrm{AA}}_{n} \gamma^{\mathrm{AA}}_{0} \left|g(\bm{k})\right| 
+ \beta^{\mathrm{AA}}_{n} \left| \gamma^{\mathrm{AA}}_{3}\left|g(\bm{k})\right| +\alpha^{\mathrm{AA}}_{n}\gamma^{\mathrm{AA}}_{1} \right|$,  with $\alpha^{\mathrm{AA}}_{n}= -1, -1, +1, +1$ and $\beta^{\mathrm{AA}}_{n} = -1, +1, -1, +1$ for the four bands, respectively \cite{Zheng2023}.
Expanding the structure factor $g(\bm{k})$ around the  $\bm{K}$ point as
$|g(\bm{K}+\delta\bm{k})|\approx\frac{3}{2}a_{0} |\delta\bm{k}|$, 
the energy becomes:
\begin{align}
		\epsilon^{\mathrm{AA}}_{n\bm{k}} = \hbar \alpha^{\mathrm{AA}}_n 
		\left(v_0^{\mathrm{AA}} 
		+  \beta^{\mathrm{AA}}_n v_3^{\mathrm{AA}}\right)
		\left|\delta \bm{k}\right| + \beta^{\mathrm{AA}}_n \gamma_{1}^{\mathrm{AA}}\,.
\end{align}
with the effective velocities defined as $v_0^{\mathrm{AA}} = \frac{3a_0\gamma_0^{\mathrm{AA}}}{2\hbar}$ and $v_3^{\mathrm{AA}} = \frac{3a_0\gamma_3^{\mathrm{AA}}}{2\hbar}$.
For comparison, the eigen-energy of monolayer graphene reads
$\epsilon^{\mathrm{Gr}}_{n\bm{k}}=\pm\hbar v_{\mathrm{F}}|\delta\bm{k}|$,
       with the Fermi velocity for monolayer graphene $v_{\mathrm{F}}=\frac{3a_0\gamma^{\mathrm{Gr}}_0}{2\hbar}$ and the nearest-neighbor hopping parameter $\gamma^{\mathrm{Gr}}_0=2.7$ eV. 
For the monolayer graphene,  the third-order nonlinear conductivity has a known analytical form that scales as the square of the Fermi velocity $\sigma^{(3)}_{\mathrm{gr}}\propto v_{\mathrm{F}}^2$ \cite{Cheng2020,Cheng2014}.
Therefore, when approximating the AA-BG response as the sum of two monolayer responses at offset chemical potentials
$\mu=\pm\gamma^{\mathrm{AA}}_1=\pm0.361$ eV, each monolayer contribution must be weighted by the square of the corresponding Fermi velocity ratio.
The scaled superposition is given by
\begin{widetext}
\begin{align}
\sigma^{(3)}_{\mathrm{AA}}=
\left(\frac{v^{\mathrm{AA}}_{0} + v^{\mathrm{AA}}_{3}}{v_{\mathrm{F}}}\right)^2
\left.\sigma_{\mathrm{gr}}^{(3)}\right|_{\mu=\gamma_1^{\mathrm{AA}}}
+
\left(\frac{v^{\mathrm{AA}}_{0} - v^{\mathrm{AA}}_{3}}{v_{\mathrm{F}}}\right)^2
\left.\sigma_{\mathrm{gr}}^{(3)}\right|_{\mu=-\gamma_1^{\mathrm{AA}}}\,.
\label{Gr-AA}
\end{align}
\end{widetext}
This result indicates that the characteristic jerk response in AA-BG originates primarily from the  intra-cone transitions,  while inter-cone contributions are negligible. 
Consequently, the response of AA-BG can be accurately approximated as a superposition of two independent monolayer-like responses, each at an effective chemical potential offset by $\pm\gamma_1^{\mathrm{AA}}$.

Similar to AA-BG,  the band structure of AAA-TG also comprises three layers of monolayer-like energy bands
 (labeled I, II, and III, represented by the blue, red, and black curves in Fig.\,\ref{fig2}\,(e), respectively).  
The central bands (set II) preserve the linear dispersion like the monolayer graphene, while the outer bands (sets I and III) are  symmetrically offset in energy by $\pm0.255$ eV. 
Different to AA-BG, AAA-TG exhibits strict electron-hole symmetry.

Figure\,\ref{fig2}\,(f) presents the jerk conductivity spectra of AAA-TG, which exhibit pronounced peaks at approximately 0.51 eV related to the single-photon transition within a set of linear bands, analogous to the mechanism observed in AA-BG. 
Furthermore, a comparative analysis also reveals good agreement between AAA-TG's conductivity spectra and the superposition of monolayer responses at $\mu=0, \pm0.255$ eV, scaled by a factor which is from the Fermi velocity differences between AAA-TG and monolayer graphene.
The low-energy bands of AAA-TG near K point can be written as
$\epsilon^{\mathrm{AAA}}_{n\bm{k}} = \alpha^{\mathrm{AAA}}_{n} \gamma^{\mathrm{AAA}}_{0} \left|g(\bm{k})\right| + \sqrt{2} \beta^{\mathrm{AAA}}_{n} \gamma^{\mathrm{AAA}}_{1}$, 
	with $\alpha^{\mathrm{AAA}}_{n}= -1, -1, -1, +1, +1, +1$ and $\beta^{\mathrm{AAA}}_{n} = -1, 0, +1, -1, 0, +1$, respectively.
	Introducing the Fermi velocity $v_0^{\mathrm{AAA}} = \frac{3a_0\gamma_0^{\mathrm{AAA}}}{2\hbar}$,
	the linearized low-energy dispersion becomes
		\begin{align}
		 \epsilon^{\mathrm{AAA}}_{n\bm{k}}  = 
		 \hbar \alpha^{\mathrm{AAA}}_{n} v_{0}^{\mathrm{AAA}} \left|\delta\bm{k}\right| 
		 + \sqrt{2} \beta^{\mathrm{AAA}}_{n} \gamma^{\mathrm{AAA}}_{1}\,,
	\end{align}
and the total jerk response of AAA-TG can be approximated as the weighted sum of three monolayer contributions
\begin{widetext}
\begin{align}
\sigma^{(3)}_{\mathrm{AAA}}=\left(
\frac{v_0^{\mathrm{AAA}}} {v_{\mathrm{F}}}
\right)^2\left(
\left.\sigma^{(3)}_{\mathrm{gr}}\right|_{\mu=\sqrt{2}\gamma^{\mathrm{AAA}}_{1}}
+\left.\sigma^{(3)}_{\mathrm{gr}}\right|_{\mu=0\,\text{eV}}
+\left.\sigma^{(3)}_{\mathrm{gr}}\right|_{\mu=-\sqrt{2}\gamma^{\mathrm{AAA}}_{1}}\right)\,.
\label{Gr-AAA}
\end{align}
\end{widetext}
It confirms that AAA-TG's jerk conductivity remains closely linked to the monolayer graphene response. 
Additionally, significant peaks appear near zero photon energy, originating from the enhanced density of states around the Dirac point, intralayer transitions in the middle layer, and the divergence of Drude-type contributions.

\subsection{AB-BG, ABA-TG and ABC-TG}
\label{Sec42}

\begin{figure*}[!htp]
	\centering
	\includegraphics[width=\linewidth]{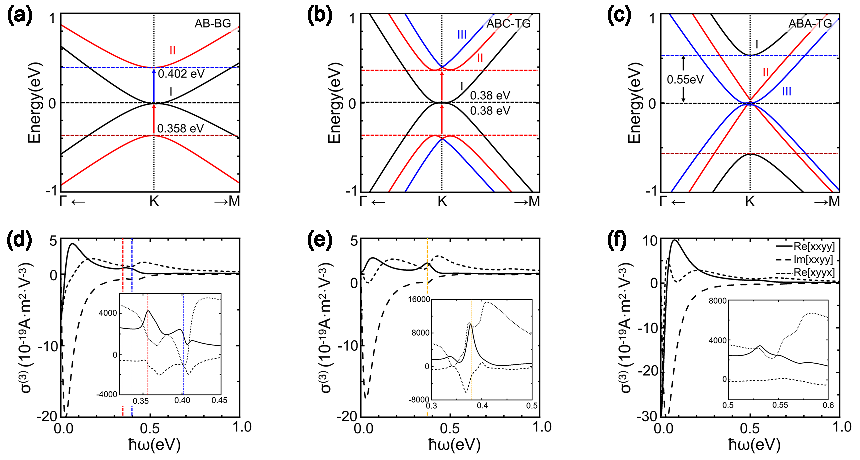}
	\caption{
	Electronic band structures of (a) AB-BG, (b) ABC-TG, and (c) ABA-TG. 
	Spectra of the jerk conductivities Re[$\sigma^{(3);xxyy}$], Im[$\sigma^{(3);xxyy}$], and Re[$\sigma^{(3);xyyx}$] at a chemical potential $\mu=0$ eV for (c) AB-BG, (e) ABC-TG, and (f) ABA-TG, with insets for small damping of $\gamma=0.5$ meV.
	}
	\label{fig3}
\end{figure*}

Then we turn to group II involving parabolic-like energy bands.
The band structures for AB-BG and ABC-TG are shown in Figs.\,\ref{fig3}\,(a) and \ref{fig3}\,(b), respectively.
AB-BG features two sets of parabolic bands (labeled I and II) that intersect near the charge neutrality point (CNP) at $-\frac{\Delta^\prime}{2} = -0.011$ eV, giving rise to low-energy electron-hole asymmetry.
ABC-TG displays a more complex band structure comprising three sets of bands (labeled I, II and III), where the bands from set I intersect at the CNP,  set II exhibits a distinctive ``Mexican-hat'' dispersion with band edges near $\pm0.38$ eV, and set III  is approximately parabolic.
Unlike AB-BG, ABC-TG preserves electron-hole symmetry, with its CNP exactly at zero energy.

These unique band structures are clearly reflected in their jerk conductivity spectra, as shown in Figs.\,\ref{fig3}\,(d) and \ref{fig3}\,(e) at $\mu=0$ eV.
In the lower-energy region below 0.1 eV, the prominent peak emerges, originating from Drude-type transitions between the two crossing bands.
At higher energies, additional spectral structures emerge, which can be attributed to interband transitions between non-crossing bands. 
For example, 
in AB-BG, 
two distinct peaks appear around 0.358 eV and 0.402 eV, 
while ABC-TG exhibits a pronounced peak near 0.38 eV.
The insets, obtained with small damping $\gamma=0.5$ meV, clearly resolve these peak positions.
Other spectral features arise from more complex interband transitions.

In contrast, ABA-TG exhibits a markedly different band structure, as discussed in Sec.\,\ref{Sec3} and reproduced in Fig.\,\ref{fig3}\,(c).
The corresponding jerk conductivity spectra are presented in Fig.\,\ref{fig3}\,(f).
We note that their line shapes are similar to those of shift conductivity, as both correspond to single-photon processes. However, their dependencies on chemical potential show distinct differences, which will be analyzed in the following section.

\subsection{Chemical potential dependence in AA-BG and AAA-TG}

\begin{figure*}[!htp]
	\centering
	\includegraphics[trim=0cm 0cm 0cm 0cm, clip, width=\linewidth]{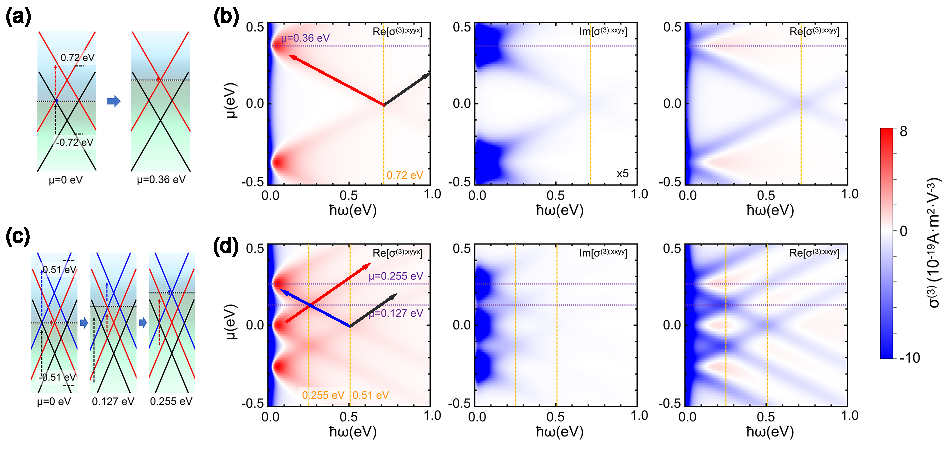}
	\caption{Schematic diagram of different chemical potentials in (a) AA-BG and (c) AAA-TG.
	Spectra of the jerk conductivities Re[$\sigma^{(3);xxyy}$], Im[$\sigma^{(3);xxyy}$], and Re[$\sigma^{(3);xyyx}$] continuously changing with chemical potential in (b) AA-BG and (d) AAA-TG.
	}
	\label{fig4}
\end{figure*}

Figures\,\ref{fig4}\,(a) and \ref{fig4}\,(b) present the schematic energy dispersion and the jerk conductivity of AA-BG, respectively,  at different chemical potentials.
As mentioned earlier, at $\mu=0$ eV, all conductivity components exhibit the peaks around 0.72 eV, originating from the intralayer transition processes. 
As $\mu$ increases (taking positive values as an example), this intralayer transition peak splits into two: one undergoes a redshift and the other a blueshift as the Fermi surface shifts toward the Dirac point K$'$.
This splitting reflects the persistence of intralayer transitions, with the redshifted peak corresponding to transitions in the upper layer (red arrow) and the blueshifted peak originating from the lower layer (black arrow).
Moreover, since the redshifted peak lies closer to the K$'$ point---where the density of states is higher---the low-energy split peak intensifies, while the high-energy counterpart diminishes.
When $\mu=0.361$ eV, the Fermi surface aligns with the Dirac point K$'$ of upper layer, leading to an anomalous divergence in the redshifted conductivity peak due to the Drude-type transition.
As the chemical potential increases further, the spectral features shift toward higher energies.
Throughout this evolution, the chemical potential dependent jerk conductivity spectra maintains mirror symmetry about the $\mu = 0$ eV, a consequence of the intrinsic electronic band symmetry of AA-BG.

Interestingly, a similar evolution of jerk conductivity emerges in AAA-TG, with distinct spectral features, as shown in Figs.\,\ref{fig4}\,(c) and \ref{fig4}\,(d).
At zero chemical potential, all jerk conductivity components exhibit dual resonance peaks. 
As the chemical potential increases (again taking positive values as an example), the higher-energy peak splits into two components, shown as one exhibiting blueshift accompanied by intensifying intensity, while the other showing gradual redshift with weakening intensity. 
In contrast, the lower-energy peak exhibits a blueshift accompanied by a reduction in intensity.
When the Fermi level reaches the first critical point at $\mu=0.127$ eV---corresponding to the band intersection between the top and middle layers---the redshifted  higher-energy branch merges with the ascending blueshifted low-energy peak at a photon energy of 0.255 eV. 
Subsequently, this merged peak will also split into two components. 
Upon Fermi level touching the second critical point $\mu= 0.255$ eV,  which aligns with the Dirac  point K$'$, the redshifted sub-branch reaches its maximal value in the low-energy region. 
Beyond this point, this peak will undergo a blueshift accompanied by a decrease in peak intensity as potential further rises.
As in the AA-BG case, these spectral  behaviors confirm that the intralayer transition plays a major role in the jerk conductivity of AAA-TG. 
The corresponding transition processes are illustrated schematically in Fig.\,\ref{fig4}\,(c).

\subsection{Chemical potential dependence in AB-BG, ABA-TG and ABC-TG}

\begin{figure*}[ht]
	\centering
	\includegraphics[width=\linewidth]{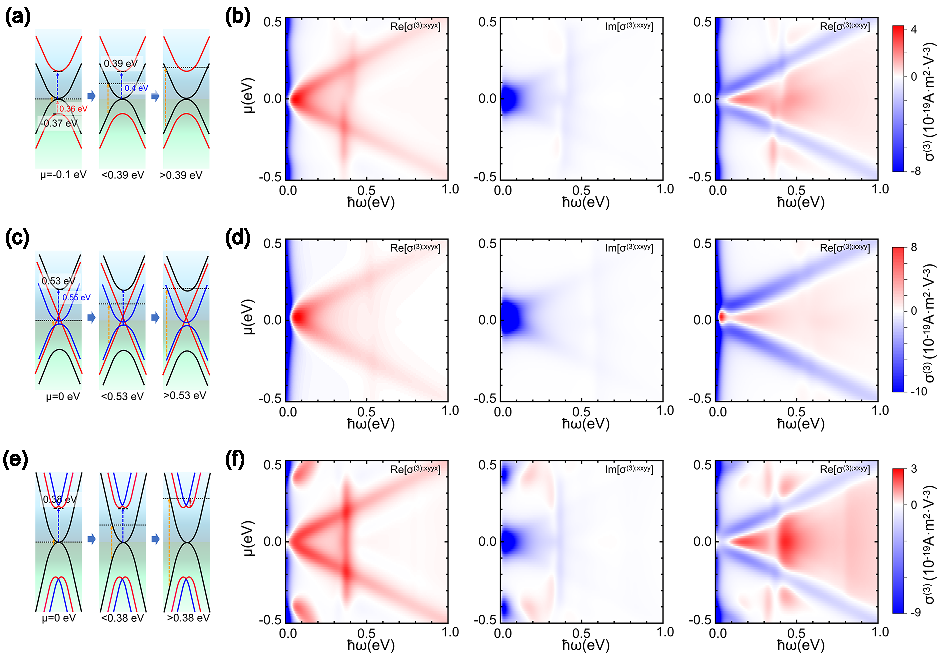}
	\caption{Schematic diagram of different chemical potentials in (a) AB-BG, (c) ABA-TG, and (e) ABC-TG.
	Spectra of the jerk conductivities Re[$\sigma^{(3);xxyy}$], Im[$\sigma^{(3);xxyy}$], and Re[$\sigma^{(3);xyyx}$] continuously changing with chemical potential in (b) AB-BG, (d) ABA-TG, and (f) ABC-TG.
	}
	\label{fig5}
\end{figure*}

Next, we investigate the chemical potential dependent jerk conductivity for group II of few-layer graphene.
Figures\,\ref{fig5}\,(a) and \ref{fig5}\,(b) show the schematic energy dispersion and the jerk conductivity spectra of AB-BG, respectively.
Owing to the intrinsic electron-hole asymmetry in AB-BG, we consider variations in chemical potential starting from its CNP, located at $\mu=-0.011$ eV.
As the Fermi surface moves away from CNP, the lower-energy conductivity peak exhibits a blueshift.
This behavior is attributed to the increase in photon energy required for interband transitions near the Fermi surface, as indicated by yellow arrows in Fig.\,\ref{fig5}\,(a).
In addition, the spectra clearly show two distinct high-energy peaks at 0.402 eV for positive $\mu$ and 0.358 eV for negative $\mu$, where positions of these two peaks remain stable against changes in the chemical potential.
It is because that they originate from two different interband transitions indicated by the blue and red arrows in Fig.\,\ref{fig5}\,(a), respectively.
As a result, the overall spectral shape appears asymmetric with respect to the CNP.

Similarly, the chemical potential dependence of the jerk conductivity in ABA-TG is presented in
Fig.\,\ref{fig5}\,(c--d).
Consistent with its shift current response, 
ABA-TG exhibits the electron-hole asymmetry. 
As the chemical potential $\mu$ deviates from its CNP, 
the  lower-energy peak, originating from interband transitions within set II and/or set III, undergoes a continuous blueshift.
This shift results from the increased energy required for vertical transitions as the Fermi level moves away from the band intersection.
However, the two higher-energy peaks, corresponding to combined interband transitions at approximately (0.53 and 0.55 eV, and 0.55 and 0.57 eV), remain essentially fixed in position.

ABC-TG exhibits significantly greater band structure complexity due to its three sets of parabolic subbands, which generate three main carrier transition pathways with distinct spectral signatures: (1) low-energy interband carrier transitions between bands of set I near the Dirac point; 
(2) high-energy transitions involving combinations of bands of set I and II or set I and III interactions; and (3) transitions between the parallel-aligned band from set II and set III (Figs.\,\ref{fig5}\,(e) and \ref{fig5}\,(f)). 
As previous report, the carrier transition (1) corresponds to the conductivity peak at low-energy region, and the peak exhibits a monotonically blue-shift trend as chemical potential increases. 
The carrier transition (2) give rise to the high-energy conductivity peak at a photon energy of 0.38 eV. Since the set I-II or set I-III are almost parallel to each other, this unchanged transition energy results in the remaining peak position with the increase of potential. 
Notably, the third transition pathway (3) between II and III becomes accessible only when chemical potential exceeds 0.38 eV, enabling a distinct conductivity peak at 0.36 eV. 
This lower-energy feature arises from the slightly smaller interband energy between sets II and III compared to I-II/I-III separations. Due to the quasi-parallel alignment of sets I and II/III, the conductivity peak preserves its position despite potential variations.

Furthermore, unlike AB-BG and ABA-TG, ABC-TG is also observed exhibiting a transition zero point precisely at $\mu=0$ eV, where its conduction and valence bands demonstrate symmetry relative to this chemical potential. 
Consequently, the material's chemical potential-dependent conductivity evolution manifests as a mirror-symmetric process, with spectral features evolving identically in both the electron-doped and hole-doped regimes.

\section{Estimations for photocurrent}
\label{Sec5}

With the shift and jerk conductivities established, we now proceed to calculate the corresponding photocurrent densities at resonant photon energies for various polarization configurations.
The incident light, as introduced in Eq.\,\eqref{field}, takes the form
\begin{align}
		\bm{E}_{\omega}=E_{\omega}\left(\cos\theta\hat{\bm{x}} +
\sin\theta e^{i\varphi}\hat{\bm{y}}\right)\,,
\end{align}
where $\theta$ denotes the polarization angle relative to the $x$-direction,  and $\varphi$ is the phase difference between the two orthogonal components.
Substituting these expressions into Eq.\,\eqref{current}, 
the shift current is given by
	\begin{align}
		\bm{J}_{\text{dc}}^{(2)} = 2\sigma^{(2);xxy}(\omega,-\omega)E_{\omega}^{2}\left(
\sin2\theta\cos\varphi\hat{\bm{x}}
+
\cos2\theta\hat{\bm{y}}\right)\,,
	\end{align}
where $J^{(2);x}_{\text{dc}}$ depends on both $\theta$ and $\varphi$, while $J^{(2);y}_{\text{dc}}$ depends only on $\theta$.
For the jerk current, the dc field is applied along the $x$-direction, i.e., $\hat{\bm{e}}_\parallel=\hat{\bm{x}}$ and  $\hat{\bm{e}}_{\bot}=\hat{\bm{y}}$), giving
\begin{align}
\bm{E}_{\mathrm{dc}} = E_{\mathrm{dc}}\hat{\bm{x}}\,.
\end{align}
Note that the specific choice of the dc field direction does not affect the generality of the results presented here, as responses for other orientations can be obtained via symmetry transformations.
	The jerk current then becomes
	\begin{widetext}
	\begin{align}
		\bm{J}_{\text{dc}}^{(3)} = &
		\left[6\sigma^{(3);xyyx}(\omega,-\omega,0)
		+
		 12 \text{Re}[\sigma^{(3);xxyy}(\omega,-\omega,0)]  \cos^2\theta  \right]E_{\omega}^{2}E_{\text{dc}} \hat{\bm{x}}
		\notag\\
		&
	+6  \left( \text{Re} [\sigma^{(3);xxyy}(\omega, -\omega, 0)]\cos\varphi - \text{Im} [\sigma^{(3);xxyy}(\omega, -\omega, 0)] \sin\varphi \right) E_{\omega }^{2} E_{\text{dc}} \sin2\theta \hat{\bm{y}}	\,.
	\end{align}
	\end{widetext}
	Here $J^{(3);x}_{\text{dc}}$ contains a polarization-independent contribution from $\sigma^{(3);xyyx}$ and a $\theta$-dependent term involving $\sigma^{(3);xxyy}$; $J^{(3);y}_{\text{dc}}$ depends on both $\theta$ and $\varphi$.
For linearly polarized light ($\varphi = 0$), the transverse ($J^{(3);y}_{\text{dc}}$) jerk current reaches its maximum at $\theta = \pi/4$ ($45^\circ$-polarization).

To evaluate the photocurrent magnitudes,  we consider linearly polarized light with $\theta=0$, $\pi/4$, $\pi/2$ and $\varphi=0$, as well as circularly polarized light with $\theta=\pi/4$ and $\varphi=\pi/2$. 
Setting $E_{\mathrm{dc}}=10^5$ V/m and $E_\omega=10^7$ V/m, the
resulting shift and jerk current values are summarized in Table\,\ref{Tab2}.
The shift current is observed exclusively in ABA-TG and only under linearly polarized illumination, yielding a self-driving current density of  18.8 A/m along $y$-direction for  $x$- and $y$-polarization, while along $x$-direction for $45^\circ$-polarization.
In contrast, the jerk current manifests universally across all bilayer and trilayer graphene systems.
It shows three distinctive characteristics: 
First, under linear polarization, both $x$- and $y$-polarized light generate a photocurrent only along the $x$-direction, but with different magnitudes. This difference arises because for $x$-polarization ($\theta = 0$), only $\sigma^{(3);xyyx}$ contributes to the current, whereas for $y$-polarization ($\theta = \pi/2$), both $\sigma^{(3);xyyx}$ and $\sigma^{(3);xxyy}$  are active.
Second, for linear polarizations other than $x$- and $y$-directions,  the jerk current direction is no longer locked to the applied dc field orientation. 
With the dc field applied along $x$, such polarizations induce both longitudinal ($J^{(3);x}_{\text{dc}}$) and transverse ($J^{(3);y}_{\text{dc}}$) components, with the latter component resembling the anomalous Hall effect reported in topological materials. 
The transverse component reaches its maximum at $\theta = \pi/4$.
Third,
the jerk current exhibits strong dependence on  the material's lattice symmetry, the static electric field, and the intensity and polarization state of the incident electromagnetic wave. 
This material-dependent anisotropy suggests potential applications in polarization-sensitive optoelectronic devices where current direction could be engineered through strategic combination of field orientations and material selection.

\begin{table*}[!htp]
	\centering
	\renewcommand\arraystretch{1.4}
	\caption{Numerical values of the shift and jerk currents for few-layer graphene under specific optical fields and photon energies with $E^x_{\mathrm{dc}}=10^5$ V/m and $E_\omega=10^7$ V/m.
	}
\begin{tabular}{ccccccccc}
\hline 
\hline 
Current & Material & $\mu$ (eV) & Polarization & $\theta$ & $\varphi$ & $\hbar\omega$ (eV) & $J_{\text{dc}}^{(n);x}$(A/m) & $J_{\text{dc}}^{(n);y}$(A/m)\tabularnewline
\hline 
\multirow{4}{*}{shift} & \multirow{4}{*}{ABA-TG} & \multirow{4}{*}{0} & \multirow{3}{*}{linear} & 0 & \multirow{3}{*}{0} & \multirow{4}{*}{0.555} & 0 & $-18.8$\tabularnewline
\cline{5-5}\cline{8-9}
 &  &  &  & $\pi/2$ &  &  & 0 & 18.8\tabularnewline
\cline{5-5}\cline{8-9}
 &  &  &  & $\pi/4$ &  &  & 18.8 & 0\tabularnewline
\cline{4-6}\cline{8-9}
 &  &  & circular & $\pi/4$ & $\pi/2$ &  & 0 & 0\tabularnewline
\hline 
\multirow{20}{*}{jerk} & \multirow{4}{*}{AA-BG} & \multirow{4}{*}{0} & \multirow{3}{*}{linear} & 0 & \multirow{3}{*}{0} & \multirow{4}{*}{0.709} & $-2.2$ & 0\tabularnewline
\cline{5-5}\cline{8-9}
 &  &  &  & $\pi/2$ &  &  & $-10.7$ & 0\tabularnewline
\cline{5-5}\cline{8-9}
 &  &  &  & $\pi/4$ &  &  & $-6.4$ & 4.3\tabularnewline
\cline{4-6}\cline{8-9}
 &  &  & circular & $\pi/4$ & $\pi/2$ &  & $-6.4$ & $-0.8$\tabularnewline
\cline{2-9}
 & \multirow{4}{*}{AB-BG} & \multirow{4}{*}{0} & \multirow{3}{*}{linear} & 0 & \multirow{3}{*}{0} & \multirow{4}{*}{0.361} & 17.6 & 0\tabularnewline
\cline{5-5}\cline{8-9}
 &  &  &  & $\pi/2$ &  &  & 7.0 & 0\tabularnewline
\cline{5-5}\cline{8-9}
 &  &  &  & $\pi/4$ &  &  & 12.3 & 5.3\tabularnewline
\cline{4-6}\cline{8-9}
 &  &  & circular & $\pi/4$ & $\pi/2$ &  & 12.3 & $-4.0$\tabularnewline
\cline{2-9}
 & \multirow{4}{*}{AAA-TG} & \multirow{4}{*}{0} & \multirow{3}{*}{linear} & 0 & \multirow{3}{*}{0} & \multirow{4}{*}{0.499} & 3.6 & 0\tabularnewline
\cline{5-5}\cline{8-9}
 &  &  &  & $\pi/2$ &  &  & $-20.7$ & 0\tabularnewline
\cline{5-5}\cline{8-9}
 &  &  &  & $\pi/4$ &  &  & $-8.5$ & 12.1\tabularnewline
\cline{4-6}\cline{8-9}
 &  &  & circular & $\pi/4$ & $\pi/2$ &  & $-8.5$ & $-2.3$\tabularnewline
\cline{2-9}
 & \multirow{4}{*}{ABA-TG} & \multirow{4}{*}{0.2} & \multirow{3}{*}{linear} & 0 & \multirow{3}{*}{0} & \multirow{4}{*}{0.372} & $-1.7$ & 0\tabularnewline
\cline{5-5}\cline{8-9}
 &  &  &  & $\pi/2$ &  &  & $-25.9$ & 0\tabularnewline
\cline{5-5}\cline{8-9}
 &  &  &  & $\pi/4$ &  &  & $-13.8$ & 12.1\tabularnewline
\cline{4-6}\cline{8-9}
 &  &  & circular & $\pi/4$ & $\pi/2$ &  & $-13.8$ & $4.0$\tabularnewline
\cline{2-9}
 & \multirow{4}{*}{ABC-TG} & \multirow{4}{*}{0} & \multirow{3}{*}{linear} & 0 & \multirow{3}{*}{0} & \multirow{4}{*}{0.371} & 24.2 & 0\tabularnewline
\cline{5-5}\cline{8-9}
 &  &  &  & $\pi/2$ &  &  & 5.8 & 0\tabularnewline
\cline{5-5}\cline{8-9}
 &  &  &  & $\pi/4$ &  &  & 15.0 & 9.2\tabularnewline
\cline{4-6}\cline{8-9}
 &  &  & circular & $\pi/4$ & $\pi/2$ &  & 15.0 & $-7.8$\tabularnewline
\hline 
\hline 
\end{tabular}
	\label{Tab2}  
\end{table*}

\section{CONCLUSION}
\label{Sec6}

This work elucidates the generation mechanism of nonlinear photogalvanic current in few-layer graphene through a tight-binding framework, establishing the critical roles of symmetry breaking and band engineering in governing shift and jerk currents. 
We demonstrate that broken inversion symmetry exclusively enables shift current generation in ABA-stacked trilayer graphene, where optimized chemical potential ($\mu=0.36$ eV) and photon energy ($\hbar\omega=0.555$ eV) yield a self-driven current density of 18.8 A$\cdot$m$^{-1}$ under a linear polarized field
$E_\omega=10^7$ V/m with peak conductivity reaching $1.21\times10^{-13}$ A$\cdot$m$^{-1}\cdot$V$^{-2}$. 
In striking contrast, the jerk currents---as third-order nonlinear phenomena---manifest universally across all stacking configurations (AA/AB bilayers; AAA/ABA/ABC trilayers), though their activation necessitates an in-plane static electric field $E_{\mathrm{dc}}^x=10^5$ V/m. 
Crucially, chemical potential tuning  enables dynamic spectral modulation from THz to near-infrared regimes, while stacking-dependent band topologies (linear Dirac cones vs. parabolic bands) dictate transition pathways and current anisotropy. 
These findings establish a universal symmetry--band--field coupling paradigm, providing foundational principles for designing programmable nonlinear optoelectronic devices based on van der Waals heterostructures.

\begin{acknowledgments}

This work has been supported by
National Natural Science Foundation of China (NSFC) (52025022, 12274065), funded from the Jilin Provincial Scientific and Technological Development Program  (20240101318JC, 20250102010JC).

\end{acknowledgments}


\bibliography{1.bib}
\end{document}